\documentclass[conference]{IEEEtran}
\let\labelindent\relax
\usepackage{enumitem}

\usepackage{blindtext, graphicx}
\usepackage{cite}

\ifCLASSINFOpdf
\else
\fi
\usepackage[cmex10]{amsmath}
\usepackage{array}

\usepackage{mdwmath}
\usepackage{mdwtab}

\usepackage{stfloats}

\usepackage{url}

\begin{document}
%
\title{Graph of Virtual Actors (GOVA): a Big Data \\ Analytics Architecture for IoT}

\author{\IEEEauthorblockN{The-Hien Dang-Ha}
\IEEEauthorblockA{The Faculty of Mathematics and Natural Sciences\\Department of Informatics\\
University of Oslo\\
1072 Blindern 0316 Oslo, Norway\\
hthdang@student.matnat.uio.no}
\and
\IEEEauthorblockN{Davide Roverso}
\IEEEauthorblockA{Analytics Department \\ eSmart Systems \\
1783 Halden, Norway \\
Davide.Roverso@esmartsystems.com}
\and
\IEEEauthorblockN{Roland Olsson}
\IEEEauthorblockA{Faculty of Computer Sciences\\ Department of Informatics \\ Hogskolen i Ostfold \\
1757 Halden, Norway \\
roland.olsson@hiof.no}}

\maketitle

\begin{abstract}
With the emergence of cloud computing and sensor technologies, Big Data analytics for the Internet of Things (IoT) has become the main force behind many innovative solutions for our society's problems. This paper provides practical explanations for the question ``why is the number of Big Data applications that succeed and have an effect on our daily life  so limited, compared with all of the solutions proposed and tested in the literature?'', with examples taken from Smart Grids. We argue that ``noninvariants'' are the most challenging issues in IoT applications, which can be easily revealed if we use the term ``invariant'' to replace the more common terms such as ``information'', ``knowledge'', or ``insight'' in any Big Data for IoT research. From our experience with developing Smart Grid applications, we produced a list of ``noninvariants'', which we believe to be the main causes of the gaps between Big Data in a laboratory and in practice in IoT applications. This paper also proposes \emph{Graph of Virtual Actors (GOVA)} as a Big Data analytics architecture for IoT applications, which not only can solve the noninvariants issues, but can also quickly scale horizontally in terms of computation, data storage, caching requirements, and programmability of the system.
\end{abstract}


\section{Introduction}
With the emergence of cloud computing and sensor technologies, Big Data analytics has become the main force behind many innovative solutions for our society's issues. The technologies for collecting, storing, and transforming data into insight have never been so accessible. With all of the new Big Data technologies such as NoSQL, Map Reduce, and Spark (just to name a few), we seem to have solutions for almost all of the Big Data challenges. But as scientists with decades of experience in the field, we constantly ask \emph{``why is the number of Big Data applications that succeed and have an effect on our daily life  so limited, compared to all of the solutions proposed and tested in the literature?''}. There must be gaps between Big Data analytics in the laboratory and in practice.

Big Data is usually characterized by its 3 main properties: \emph{Volume}, \emph{Velocity}, and \emph{Variety}. These well-known 3Vs were first introduced by Doug Laney in 2001 \cite{laney2001} and then popularized by Gartner in their definition of Big Data in 2012 \cite{gartner}. Since not all Big Data applications possess all of those three properties, the gaps between Big Data in laboratory and practice vary from application to application. For example, Big Data applications in bioinformatics have to deal mainly with the big volume problem without worrying too much about velocity or variety. In these requests, the path from laboratory to practice is usually comparatively short. Building a model that can accurately predict the optimal cancer treatment strategy from a person's genetic makeup is almost guaranteed to make the impact on human society. However, in IoT applications like Smart Grid, the main challenges come from velocity and variety data characteristics. These applications require insights extracted from a large number of entities, which connect with each other through multiple layers and types of relationships. The insights need to be delivered on time or become virtually meaningless. Moreover, in many situations, taking actions based on the acquired insights can change the data context and/or the underlying model altogether. Therefore, in IoT field, developing a highly accurate model in laboratory condition hardly guarantee its usefulness in real life.

Working at a company that provides Smart Grid solutions for multiple distribution system operators in Norway, we deeply understand how challenging it is to bridge those gaps. In this paper, we suggest using the term ``invariant'' to replace more generic terms such as ``information'', ``knowledge'', or ``insight'' in Big Data for IoT research. In many situations, it could expose the nature of knowledge and help both the speaker and the audience be consciously aware of the ``noninvariant'' issue. This paper shows that most of the lab-practice gaps in Smart Grid (or IoT applications in general) are caused by the so-called ``noninvariants'', which is usually referred to as the \emph{taking data out of context} issues. From our experience on developing Smart Grid application, we produced a list of popular noninvariants and proposed graph of virtual actors (GoVA) as a Big Data solution for these issues. Moreover, the GoVA analytics architecture is not only horizontally scalable in term of computational, data storage, and caching requirements, but also in term of programmability of the system. GoVA brings the object-oriented programming (OOP) paradigm back to the system level and promotes the micro-services design to maintain the programmability of an ever-increasing complex distributed system.

Note that this paper focuses on the Big Data analytics aspect of IoT, where data is collected, processed, and analyzed at one or multiple data centers (on-premises or in the cloud).  In other words, we limit ourselves to the IoT centralized designs, where no decision is made locally on remote devices or sensors. Their roles are capped at collecting and communicating accurate data to centralized collection points, back from which they also receive commands.

The rest of the paper is structured as follows: Section \ref{sec:related_works} presents an overview of related research in the literature, Section \ref{sec:non_invariants} provides a list of common noninvaraints in IoT applications, Section \ref{sec:GoVA} explains GoVA analytics architectures in detail, and Section \ref{sec:conclude} concludes the paper.

\section{Related Work}
\label{sec:related_works}
At the Gartner Business Intelligence \& Analytics Summit 2015, Gartner predicted that through 2017, $60$ percent of big data projects will fail to go beyond piloting and experimentation, and will be abandoned \cite{gartner2015culture}. They claimed that old company mindsets and culture are the main obstacles. Some other parties tell the same story. In 2010, the MIT Sloan Management Review, in collaboration with the IBM Institute for Business Value, conducted a survey of $3000$ business executives, managers, and analysts working across more than $30$ industries \cite{lavalle2011big}. The study shows that technical issues are more often than not considered \emph{minor}, compared to other business-related issues like managerial and cultural ones. This goes in the fields where Big Data technologies are required to improve the existing business model, by revealing insights from the data which was previously too costly to store or process. However, in IoT applications where virtually all of the potential products and services are based on Big Data analytics, technical issues are the main causes of the high adoption barriers \cite{boyd2012critical, fan2013mining}. This paper goes further in arguing that \emph{noninvariants} are the answer to the question ``why is the number of Big Data applications that succeed and have an effect on our daily life is so limited?''.

Noninvariants issues are common in academia under various names such as \emph{taking data out of context} or \emph{sampling bias}. In 2005, John P. \cite{ioannidis2005most} asked and attempted to answer the question: ``why most published research findings are false'', which is similar to the theme of this paper. The author showed that claimed research findings might often be accurate measures of the \emph{prevailing bias}. The similar problem in Big Data was discussed in \cite{boyd2012critical}, where the authors argued that ``taken out of context, Big Data loses its meaning and value''. In statistics, this problem is well-known under the name \emph{sampling bias}, which cannot be completely eliminated as stated by Good et.al. 2012 \cite{good2012common}: ``With careful and prolonged planning, we may reduce or eliminate many potential sources of bias, but seldom will we be able to eliminate all of them.'' Furthermore, unlike error related to random variability coming from the nature of the phenomenon, the sampling bias cannot be assessed without external knowledge of the world \cite{weisberg2011bias}. In other words, it is impossible for a system to know from its collected data whether that data is biased or not.

In Big Data, this problem becomes even worse, especially in IoT applications. With the unprecedented scale, speed, and range of data available from IoT technologies, the phenomena that we want to analyze, model, and predict are inherently more complex. Most of these phenomena are more dynamic than what usually reflected by the data that used to explain them, which causes the fails of Big Data in practice. Hence we proposed the term \emph{noninvariant} to denote this type of problem. In the book Large Scale Inference \cite{efron2012large}, Efron provided various examples to help identify symptoms of the noninvariant problem in biomedical field (although he did not use the term noninvariant). Section \ref{sec:non_invariants} plays a similar role in the IoT field, in which we defined the term noninvariant in detail and presented various typical noninvariant examples taken from Smart Grid.

Although from a physical point of view, we can never eliminate noninvariants from any form of analytics in a partially observable environment, we can mitigate the problem by carefully designing an analytics architecture. In this paper, we introduce Graph of Virtual Actors (GoVA) as a unified Big Data analytics architecture for IoT. It combines the three following principles: virtual entities, micro-services, and graph database. GoVA architecture can mitigate the noninvariant issues in Big Data, have an adaptive caching mechanism, and be able to scale in term of computational power, data storage, and system's programmability. This section reviews some relevant work based on which we design the GoVA architecture.

\begin{figure}
    \centering
    \includegraphics[width=1\linewidth, trim={4cm 19cm 3.5cm 3cm},clip]{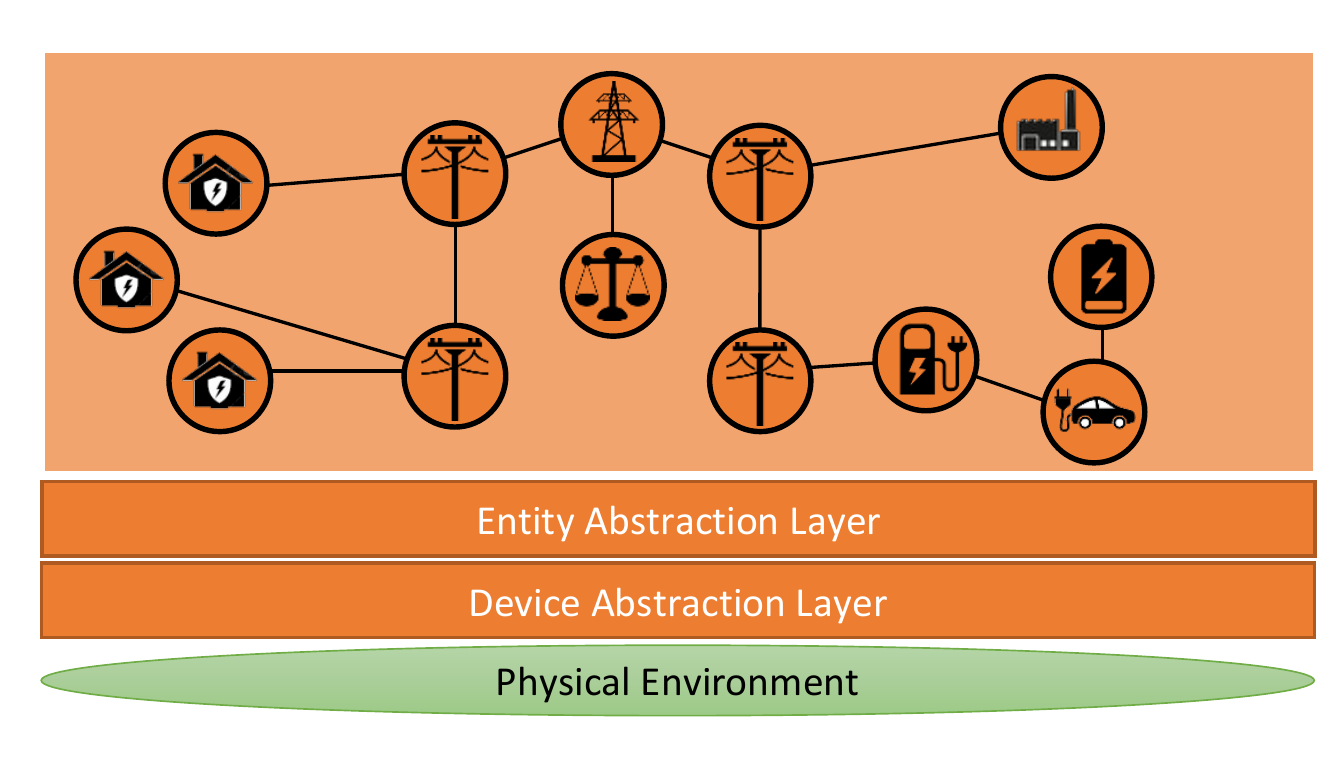}
    \caption{A simple virtual entity architecture proposed by Popovici et.al \cite{popovicicapturing}. The two abstraction layers (Device Abstraction Layer and Entity Abstraction Layer) are used to overcome the heterogeneity and complication of individual physical entities by mapping them to simple generic models. The physical environment of this example is a smart distribution grid, which displayed in detail in Figure \ref{fig:GoVA}}
    \label{fig:VirtualEntity}
\end{figure}

In IoT-related research, the idea of \emph{virtual entities} (i.e. \emph{digital artefacts} or \emph{cyber-twins}) to represent \emph{physical entities} is very natural and has been proposed multiple times \cite{walewski2013project, gazis2013unified, strohbach2015towards, lee2008cyber}. The most remarkable research is the three-year European Lighthouse Integrated Project called IoT-A (Internet-of-Things Architecture) \cite{walewski2013project}. It provides an architectural reference model in which the concept of virtual entities is one of the fundamental building blocks. IoT-A's architectural model defines virtual entities as \emph{synchronized representations} of a given set of aspects (or properties) of the physical entity. This means that relevant status changes in the physical entity are reflected by corresponding changes in its associated virtual entities and vice-versa. However, this is an abstract reference model which they do not provide any means to realize. Figure \ref{fig:VirtualEntity} illustrates a simple virtual entity architecture.

To mitigate the noninvariant issues, the analytics platform must be able to analyze relevant historical and real-time contextual conditions simultaneously. However, this is still considered as an unsolved problem in Big Data literature \cite{fan2013mining}. It requires the analytics architecture to support a \emph{multi-temperature} data management where frequently accessed data stored on fast storage (hot data), less-frequently accessed data stored on slightly slower storage (warm data), and rarely accessed data stored on the slowest storage an organization has (cold data) \cite{ibm2012temperature}. The challenge lays on the fact that data temperature changes differently in different part of an IoT system, each requires different caching mechanism. This dynamic requires the system to be able to adapt its caching mechanism according to the relevance of context information for certain purpose.

\begin{figure}
    \centering
    \includegraphics[width=1\linewidth, trim={3cm 15.4cm 3.8cm 5.5cm},clip]{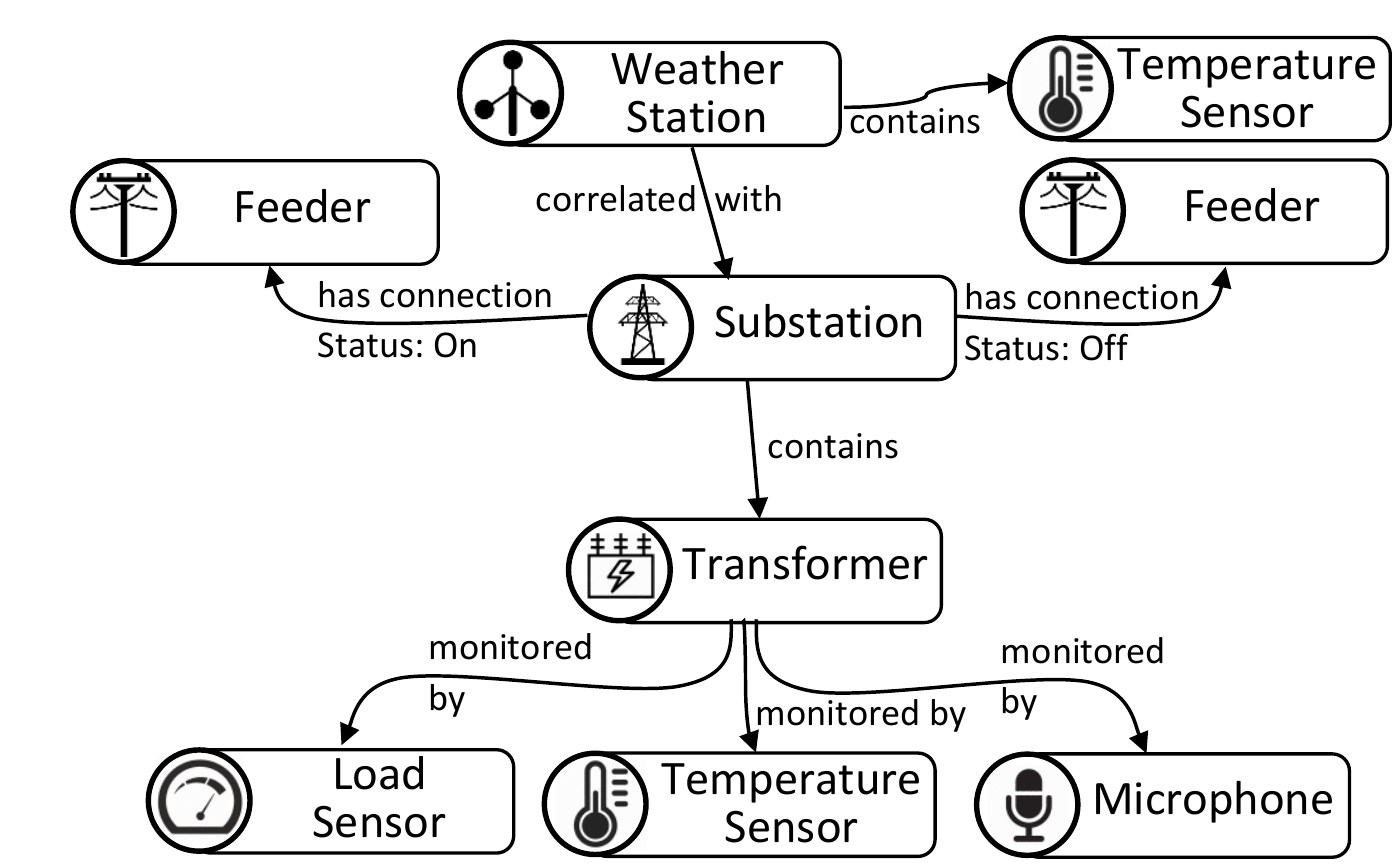}
    \caption{A small typical graph in a smart distribution system that shows the usefulness of graph database for IoT. Graph database has flexible schema, semantically represents its structure, and serves graph-based queries efficiently}
    \label{fig:GraphDb}
\end{figure}

\emph{Graph database} has been proposed as a suitable data modeling tool for applications where relations between data are as important as the data itself \cite{miller2013graph}. In Big Data analytics for IoT, subgraph matching, graph traversal, or graph analysis are commonly needed. These graph-based queries are expensive in the traditional database due to the need of recursive JOINs. A native graph database engine such as Neo4j \cite{miller2013graph}, GraphX \cite{xin2013graphx}, or Trinity\cite{shao2013trinity} can easily outperform an RDBMS in those tasks. Moreover, graph database provides a more flexible data model, where adding a new type of entity or relationship does not necessarily require a change in the database schema. Furthermore, data is stored so that it semantically represents its structure. All of these properties make graph database a useful data modeling tool for IoT. Figure \ref{fig:GraphDb} shows a small typical graph in a smart distribution system grid.

\begin{figure}
    \centering
    \includegraphics[width=0.95\linewidth, trim={1.5cm 6.5cm 5.7cm 10cm},clip]{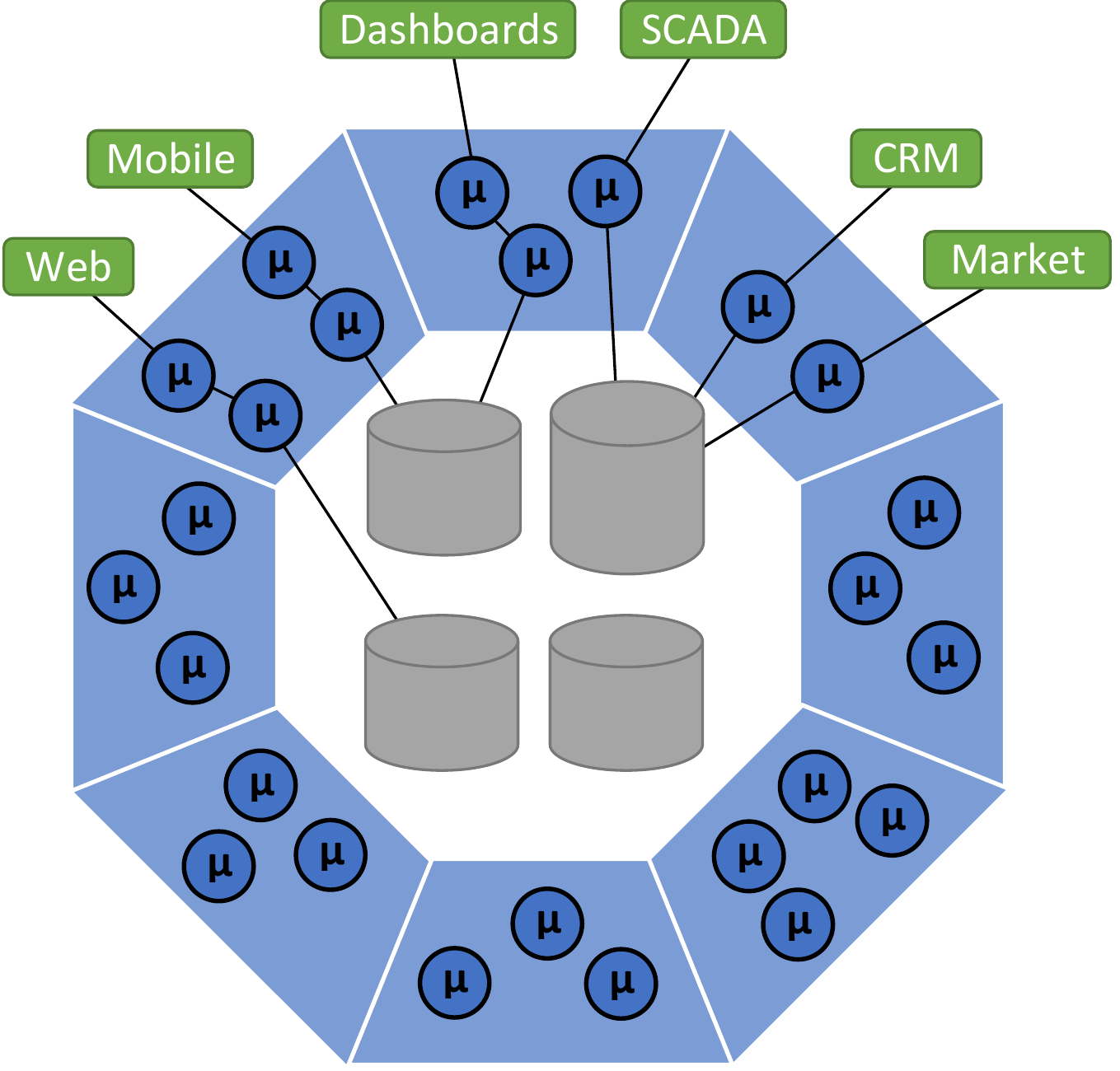}
    \caption{A micro-services architecture example. The whole system is functionally decomposed into a set of collaborating services. Each service (small micro-circle) is decoupled from other services, has a small set of responsibilities, and can be independently deployed by fully automated deployment machinery}
    \label{fig:MicroArch}
\end{figure}

\emph{Scalability} has always been a requirement for any IoT platform. Besides the ability to scale horizontally in term of computational and data storage capacity, another dimension which is equally essential but rarely mentioned in the literature is the \emph{programming scalability}. This ability is to have the team of developers expanded horizontally easily so that a newcomer can quickly understand and contribute to the project. This allows the development team to grow up in size easily. Among others, \emph{micro-services architecture} is one of the most common solutions for this scalability challenge, reportedly being used by most of the large scale websites such as Netflix \cite{netflix2016}, eBay \cite{ebay2006}, and Amazon\cite{amazon2007}\dots  In this architecture, the whole system is functionally decomposed into a set of collaborating services. Each service is decoupled from other services, has a small set of responsibilities, and can be independently deployed by fully automated deployment machinery \cite{chris2015microservices}. An example of this architecture is illustrated in figure \ref{fig:MicroArch}.

\section{The Noninvariant Challenges in Big Data Analytics for IoT Applications}
\label{sec:non_invariants}
The ultimate goal of analytics is to find the \emph{invariants} in the observed phenomenon or system. They could be invariants in behaviors of a particular component, invariants in relationships between components, or invariants in some underlying distributions. In this paper, we use the term ``invariant'' as a more explicit representation of the other generic and implicit words such as information, knowledge, or insights. This usage can reveal the challenge in bringing Big Data analytics to real life, in which the phenomenon is usually more dynamic than what is shown in the data being used to model it. 

 When extracting knowledge about a particular phenomenon, a scientist like a biologist or a chemist would take the objects of interest out of the environment and study them in laboratory (i.e., controlled) conditions. Data scientists follow a similar methodology. We get the required data from a running system, extracting, and transforming them until a good model comes out. We organize competitions and introduce various benchmarks in an attempt to find the best model for the data at hand. As a consequence, a good model developed in controlled environments is not guaranteed to work well in practice (i.e., uncontrolled environments). This happens commonly in IoT applications, where taking data out of a running system usually introduces misguided invariants. This paper uses \emph{the term noninvariants to indicate patterns/structures/properties that appear to be invariant in the data, but actually dynamic in real life.}
 
 
 In this section, we defined four different types of noninvariants that we frequently encounter in the context of Smart Grid, including context noninvariant, relationship noninvariant, identity noninvariant, and behavior noninvariant. One can easily find examples of these noninvariants in other IoT applications or Big Data analytics in general.

\subsection{Context Noninvariant}
This problem is discussed shortly in \cite{boyd2012critical}, where the authors argued that ``taken out of context, Big Data loses its meaning and value''. They provided an example where communication patterns between two persons such as social media interactions or cell coordinates, are not representative for the importance of their relationship. Spending more time with colleagues than your spouse does not necessarily imply that the ``tie strength'' of that relationship is stronger. However, the authors failed to clarify the meaning of the term ``context'', which obscures the value of their conclusion. In this paper, we redefined the term as follows, which adapted from the definition given in \cite{abowd1999towards} to be more appropriate for Big Data applications:

\begin{quote}
Context is any information that can be used to characterize the situation of entities (i.e. whether a person, place, or object) that are considered relevant to the model of the phenomenon of interest.
\end{quote}
Hence the action of \emph{taking data out of context} is equivalent to \emph{ignoring relevant contextual information} when modeling a phenomenon. This leads to the notion of \emph{context noninvariant}, where relevant contextual conditions vary at different stages of analytics, such as between the training and predicting stages. Note that the term ``context information'' used in this paper refers to all data generated by processing raw data, which could be aggregated data, feature data extracted for modeling, or model's predictions \dots

One can argue that the definition of context noninvariant issue already includes the notion of the three other types of noninvariants. This is true, since relationship, identity, and behavior of entity are context information in general. However, as you will see later, they deserve their own names due to their importance, popularity, and unique characteristic in Big Data analytics.

Due to the ``garbage in, garbage out'' principle, it is apparent that to model a phenomenon effectively, its relevant contextual conditions must be identified and monitored in the first place. This requires the analysts to have domain knowledge about the event. This problem is too wide to be captured in this paper. Instead, we want to show that in many cases, relevant contextual conditions are ignored even when its data is available in the IoT system. This is usually because some contextual conditions are challenging to handle. Load forecasting in Smart Grid can thoroughly illustrate this point. 

Load forecasting is one of the most fundamental applications of Big Data in Smart Grid, which provides essential input to other applications such as \emph{Demand Response} (DR), \emph{topology optimization} or \emph{abnormally detection} \cite{dang2015role}. Due to its crucial role in power systems, the problem has been studied by a significant amount of literature for many years \cite{taohong2015whitepaper}. In fact, there is a competition called \emph{Global energy forecasting competition} (GEFCom), which takes place every second year since $2012$ \cite{hong2014global}. The competition was about (among other tasks) predicting the hourly electricity consumption at a dozen of substations, given three years historical consumption and hourly temperature collected at nearby weather stations. This competition is interesting because it depicts the controlled environments under which data scientists usually work. In this case, we ignore the dynamic of the grid topology, by which simply changing a switch can completely change the underlying consumption models. In other words, the context when the data is collected and when the model is used to forecast can be very different, hence the name context noninvariant. Any invariant learned from this setting is at high risk of being wrong in practice.

To solve this problem, an IoT analytics platform must be able to store and return the topology information on time. Moreover, whenever there is a change in topology, the historical consumption time-series of all affected substations must be updated accordingly. This involves re-aggregating past consumptions of all consumers that connect to the substation. This topology change is a concrete example of the context noninvariant issue. Section \ref{sec:GoVA} introduces and explains why GoVA architecture can solve this type of problem.

\subsection{Relationship Noninvariant} 
Another noninvariant that the load forecasting task can point out is the \emph{relationship noninvariant}. The term refers to the dynamics of relationships between entities, which its model fails to capture. This noninvariant is severe in the Big Data era, where we try to encapsulate as much related information about the modeling target as possible. Building a predictive model of an IoT entity's behaviors usually involves finding the correlation between its behaviors and others' that it has a relationship with. Therefore, identifying and keeping track of every change in relationships between entities in an IoT system are crucial. The relationships between entities can be physical or virtual links and should be considered as temporary. For example in load forecasting, outdoor temperature is a very reliable indicator, which can predict consumption up to $70\%$ accuracy \cite{taohong2015whitepaper}. However, this is only true when the temperature is measured near the location of consumption. This so called ``nearby relationship'' is not entirely determined by physical distance. Due to the terrain of the region, (e.g., mountain, river,\dots), its consumption might not correlate best with temperature measured at the weather station with the closest distance. This relationship can vary over time, due to seasonal change, or introduction of a new weather station. By getting data out of a running system like in the GEFCom competition, we assume that these relationships are fixed, which is unlikely to be true over an extended period (in this case it is $3$ years). To overcome this issue, it requires an IoT analytics platform to have the capacity of storing, monitoring and querying millions of relationships effectively.

\subsection{Identity Noninvariant}
In \cite{jonas2006identity} and \cite{Jonas2010}, Jonas Jeff, Chief Scientist at IBM, described  \emph{identity resolution} as the single most fundamental ability of any big data \emph{sensemaking} system. Identity resolution is the ability of a system to recognize an entity of interest through different (sometimes fuzzy) ways of references. After all, we cannot build a model of an entity if we cannot correctly recognize which data is describing it. A music recommendation system is a good example. It cannot model a user's taste correctly if multiple users share the same account while account information is the only identity mechanism. Similarly, in Smart Grid, consumption pattern of a household can change completely when its owner is changed. We classified this as the \emph{identity noninvariant} issue.

The most critical identity noninvariant arises in IoT applications when the identity of a physical entity changes over time. For example, an IoT system must be able to keep every time-series of sensor data consistent with its metadata (e.g., deployment location, sensor configuration\dots). In an evolving system environment, the functional dependencies between sensors and metadata entries cannot be assumed. This issue is clearly stated in \cite{strohbach2015towards}, where the authors recommend storing each sensor value with all related metadata to ensure that later analysis can always be done in the correct context during measurement time. They demonstrated the issue by showing the need to move the same hardware to multiple locations as their system use-cases were expanded. However, they did not suggest any mechanism to efficiently store the metadata, which builds up quickly when the number of sensors increases. The GoVA proposed in this paper overcomes this problem by projecting real-world entities into the digital world. Each virtual actor in the system takes responsibility to store and retrieve all of its sensor values, which maintains the measurement context at all time.


\subsection{Behavior Noninvariant Caused by Optimization Process}
The last and also the most challenging noninvariant is the \emph{behavior noninvariant} that happens when an optimization process is carried out. This issue is well-known in game theory where an action of an actor can significantly change its environment, which in turn can alter the behaviors of other agents. Surprisingly, this issue is rarely mentioned in IoT and Smart Grid literature.

The grid topology optimization process is a good example. In this application, the network topology is optimized by dynamically controlling its switches based on consumption patterns of different regions. The general objective of this process is to balance the load, improve reliability, and extend the components' lifetime. However, first of all, this optimization process can change the measurement context of many sensors along the grid, which causes the \emph{context noninvariant} issue as mentioned earlier. Furthermore, it changes the environment in which many other actors live (e.g., transformers, charging stations, building\dots), which can potentially affect their behaviors.

Another example is the demand response (DR) application, which is one of the most fundamental features of a Smart Grid. DR is defined in \cite{qdr2006benefits} as changes in electric usage by end-use customers from their \emph{normal consumption patterns} to induce lower electricity use at times of high wholesale market prices or when system reliability is jeopardized. DR is achieved by changing consumption behaviors of different type of flexible consumers, such as dispatchable load, storage, or distributed generation. However, executing DR plan for a long enough period can change the notion of \emph{normal consumption pattern} of each consumer.

Although the solution for this noninvariant is beyond the scope of this paper, we argued that to overcome  this problem, a Smart Grid platform must be at least able to keep track of all the actions that have been done to the system (in this case the switch configurations or the DR commands), so that later analysis can be done correctly.

\section{Graph of Virtual Actors}
\label{sec:GoVA} 
Researchers and companies are working hard towards the notion of a complete end-to-end Big Data Analytics platform for IoT, which is capable of quickly digesting daunting volumes of data and producing accurate insights, predictions, and optimization plans without direct human involvement. In the previous section, we have revealed $4$ types of noninvariant issues that one must overcome first to realize such an end to end analytics system. Although there is no unique solution for all of these noninvariants, there exists a set of functional requirements that an IoT system must be able to fulfill. These requirements are listed in the first part of this section, based on which GoVA analytics architecture was designed. The GoVA design is given in the second part of this section.
\subsection{Functional Requirements}
\label{sec:requirements}
 \subsubsection{Context information management} as shown earlier, data is meaningless without the ability to retrieve all of its \emph{relevant} context information. For example, in consumption forecasting task, a consumption data point like ``consumption: 110'' means nothing if it is provided without all the relevant context information such as consumer identity, time of use, the metric being used, the weather condition, or even the grid topology at the time, etc. An IoT system must provide an efficient and effective context management mechanism, so that incoming context information can be quickly stored (implicitly or explicitly), distributed (internally or externally), and queried (on demand or schedule). This solution involves managing the relevance property of context information since this determines where, when, and how the context information is required. The relevance property of context information depends on many factors, which characterizes into the three following dimensions.
 
 \begin{description}[align=left]
        \item[Scope of relevance:] certain context information can be relevant only for a particular set of applications or system functions. For example, in a Smart Grid, consumption of a lighting device at a household is only relevant for smart building application. This information is rarely needed for demand response, topology optimization, or smart charging functions of Smart Grid. This scope of relevance determines where context information should be stored in a distributed system. \\
        
        \item[Period of relevance:] context information can become irrelevant in a matter of seconds up to the lifetime of a physical entity. For example, profiled information like the measurement unit of a sensor is static and remains valid for prolonged periods of time. Other context information like relationships between entities or sensor measurements is much more volatile, which become irrelevant more quickly. This difference suggests the need for different mechanisms for storing context information. Nonvolatile context like entity profiles can be stored implicitly in the programming structure of the application to improve efficiency and performance. In contrast, more volatile context relationship between entities must be stored explicitly, so that full control can be achieved. Since data loses value over time, the system should be able to actively compress, downsample, or delete data depending on its period of relevance.\\
        
        \item[Intensity of relevance:] some context information is requested far more frequently than others. This difference leads to the notion of \emph{temperature of data} as we mentioned before. Since a different type of context information has a different pattern on how its temperature changes, an IoT system needs to design its caching mechanism accordingly. For example, feature data extracted for some analytics tasks can be cached on fast storage, stored on slow storage, or dismissed and recomputed later based on its intensity of relevance. This variety in caching requirements makes traditional caching mechanism ineffective. 
    \end{description}
    The scope, period and intensity of relevance determine where, when, and how a particular context information should be stored. 
\subsubsection{Context Modeling} 
Context modeling is a process by which a data structure or a language is used to represent the semantic of available context information. The chosen data structure or language defines how data is represented in a context information repository. For example, the simplest context model is the key-value model, which represents context information as key-value pairs. Several other traditional context modeling techniques include markup scheme tagged encoding (e.g., xml), graphical based modeling (e.g. relational or native graph database), and ontology based modeling (e.g., RDF triplestore, web ontology language -- OWL). A good context model for IoT applications must be able to represent the structures and relationships between contexts. Moreover, an IoT context model must be flexible so that new types of contexts and relationships between them can be added easily. Information retrieval process must be simple, efficient, and provide suitable data retrieval mechanism for common querying tasks.

\subsubsection{Scalability }
In many cases, the growth of an IoT system is very unpredictable. After one contract, a pilot Smart Grid system consisting of a couple of thousands of sensors might need to scale up hundred times to embrace millions of new remote devices in a matter of months or even weeks. Without a careful design, this mission would be truly impossible. The solution requires the system to be capable of scaling horizontally, by which an increase in the number of servers can lead to a linear or sublinear boost in general performance. Furthermore, new applications are usually added to an IoT system in an incremental manner. This expansion requires human resources such as the number of developers or testers to be scalable. To achieve this, newcomers should be able to understand and start being productive quickly. This requires the system's programmability to be maintained while new applications keep being added to the system.

\subsection{Graph of Virtual Actors (GoVA)}
\begin{figure*}[htp]
        \centering
        \includegraphics[width=0.55\linewidth, trim={2.5cm 1.5cm 0.5cm 5.cm},clip]{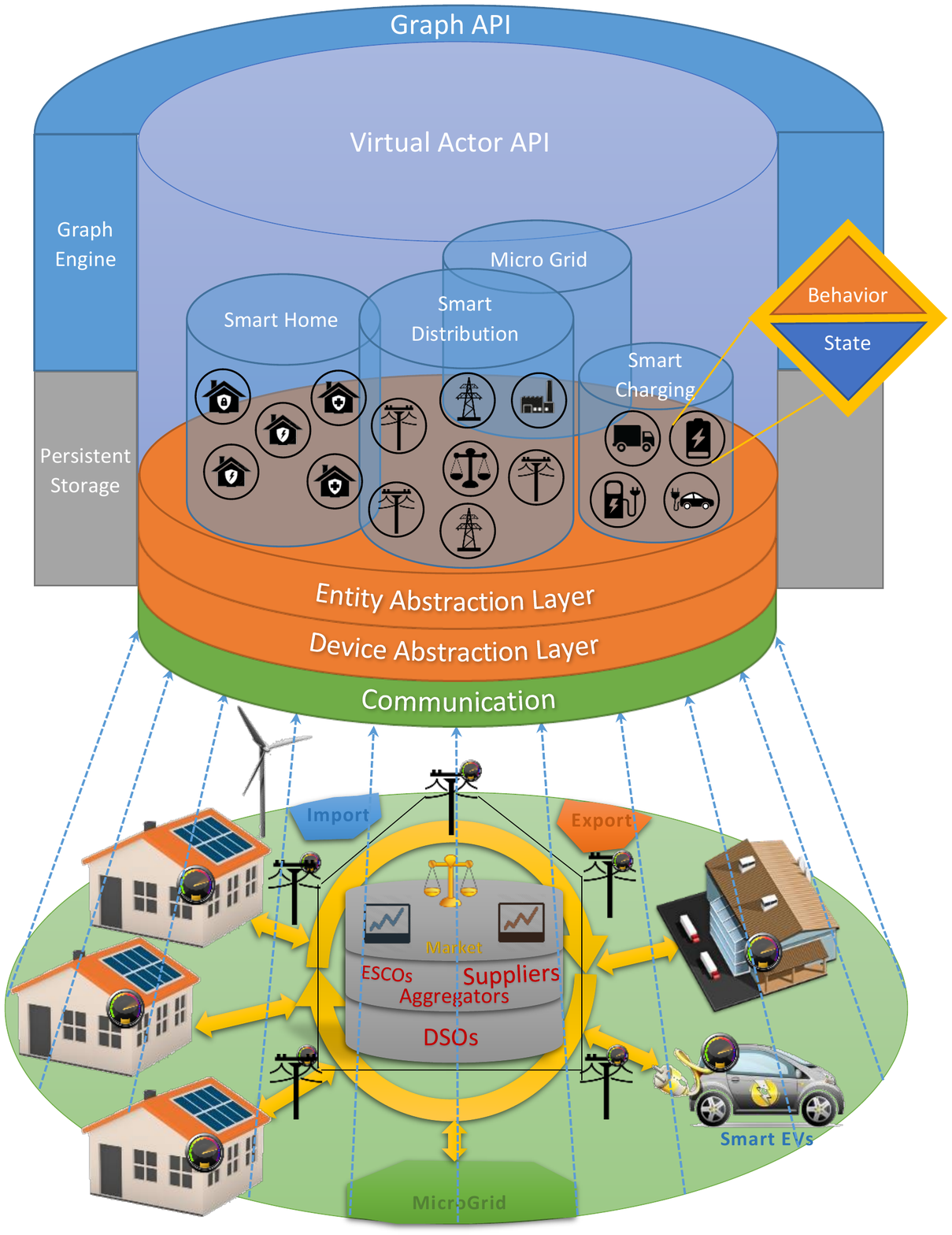}
        \caption{The general design of the Graph of Virtual Actor analytics architecture. The design combines virtual entities, micro-services architecture, and graph database. In the figure, one black circle is a virtual actor, which has its state and behavior and plays the role of a micro-service. Virtual actors are deployed to servers (i.e. silos) in a way that minimizes cross-server messaging. All virtual actors have access to a persistent global storage and a graph database. The graph database is where all the semantic and structural relationships between virtual actors are stored. Physical entities are projected into digital world through several abstraction layers which are implemented through OOP's polymorphism. The physical environment shown in this figure is a smart distribution grid (refer to \cite{dang2015role} for more detail).}
        \label{fig:GoVA}
    \end{figure*}

    In an attempt to fulfill the three above requirements, we propose the GoVA architecture, which combines the three modern architecture design principles: virtual entities, micro-services architecture, and graph database. As mentioned in the Related Work section, these three ideas have been proposed to partially solve different aspects of challenges in Big Data analytics for IoT. This paper is the first attempt to combine these three ideas to create a unified analytics architecture solution. Figure \ref{fig:GoVA} shows the general design of the GoVA architecture.

    A virtual actor (VA) in this architecture is essentially a virtual entity, which is a synchronized representation of a given set of aspects of a physical entity (or a set of physical entities). To overcome the heterogeneity and complication of individual physical entities (e.g., in protocol, networking, or functionality), one or more abstraction layers could be added. These abstraction layers are implemented through the OOP paradigm, in which polymorphism can effectively replace all conditionals or enumeration variables. This is supported by most of the available virtual actor programming framework, such as Orleans, Azure Reliable Actors, or Akka. 

Besides being a virtual entity, a VA in GoVA also plays the role of a micro-service. Each VA has a separate state and behavior, communicate with others through asynchronous message passing, and can be deployed independently with other VAs. This design is the advantage of micro-services architecture, which helps maintain the programmability of the whole IoT platform when new applications emerge. New developers only have to understand behaviors of a small subset of the VAs to be able to develop and deploy new VAs. Furthermore, each virtual actor can subscribe others to be notified when their context information changes. This ability makes the architecture proactive, event-driven, and resilient. For example, when the grid topology changes due to change in the context of some VAs (e.g., switch configuration), all affected virtual actors will be notified automatically, who then update their context information (including historical one) accordingly.

The context information management is done at the micro-service level. Each VA manages its own context information. Depending on the relevance property of its context information, a VA decides which information to keep in memory (as its state), which to store on disk, and which should be compressed or subsampled. Moreover, most VA programming frameworks allow storing deactivated VA on disk if it is not called for extended time. This approach is superior compared to other traditional global caching mechanisms since it adapts the mechanism to the specific relevance property of a particular context information. With this approach, relevant historical data can be cached as part of the state of the virtual actor, which enables handling historical and real-time data at the same time.

All VAs have access to the same persistent storage. This persistent storage can be a mix of blob storage, relational database, and NoSQL database that VAs can choose depending on their needs. This approach is feasible thanks to advance in networks. Many experiments have shown that for a typical hardware setup, reading from local disk is only about 8\% faster than reading from the disk of another node in a data center \cite{ananthanarayanan2011disk}. As shown in table \ref{tab:latency}, the round-trip latency (.5ms) is much lower than the disk seek operation (10ms) that storage can be disaggregated and distributed anywhere in the data center without significant performance degradation. Even if SSDs or NVRAM drives are used, the networking latency is still negligible if 10 Gbps network is used.    

Although \emph{disk-locality} has already become irrelevant in a data center, \emph{memory-locality} still plays an important role, since reading from local memory is two orders of magnitude faster than reading from remote host’s memory (table \ref{tab:latency}). Therefore, the system must deploy VAs in a way that the number of cross-server messages is small. This is where the graph database comes into play. In GoVA architecture, the graph database must be stored in an in-memory distributed graph engine such as Trinity or Spark GraphX to ensure its scalability. The graph database is where all the semantic and structural relationships between virtual actors are stored. Based on this information, graph partition algorithms can be employed to optimize the VAs deployment by minimizing the number of cross-sever relationships. Since not all relationships mean the same level of communication need, only some types of relationship should be considered, such as in-the-same-application or geographic closeness. 

VAs also use graph database to search for the right service providers. For example, a substation VA might want to search for a nearby weather station VA that has a temperature sensor that supports at least hourly sampling frequency. As shown before, the graph-based queries like this can only be done effectively on a native graph database. Also, graph database has a flexible schema that new types of VAs and relationships between them can be added easily.

In IoT applications where data needs to be shared and reused across enterprise or community boundaries, a formal ontology for actors, their relationships, and their services can be defined. The ontology helps external agents know how to search for the service the want in our IoT platform. It can be thought of as a schema for our graph database, which defines what type of VA exists and what kind of relationship between them is allowed. However, this approach will negatively affect the flexibility of the graph database. Moreover, data consistency needs to be maintained at all time, which can significantly affect the system's overall performance. Therefore, this is left as an option in GoVA. Note that graph database can store the ontology data itself.

    \begin{table}[tbh]
        \caption{Latency table according to Jeff Dean 2009 \cite{dean2009designs}}
        \label{tab:latency}

    \centering
    \begin{tabular}{ l  r r }
        Operation                                & Latency (ns) & (ms)\\ \hline
        L1 cache reference                       & 0.5 ns  & \\
        Branch mispredict                        & 5 ns  & \\
        L2 cache reference                       & 7 ns  & \\         
        Main memory reference                    &100 ns  & \\         
        Compress 1K bytes with Zippy             &3,000 ns & \\
        Send 1K bytes over 1 Gbps network        &10,000 ns & 0.01 ms\\
        Read 4K randomly from SSD                &150,000 ns & 0.15 ms  \\
        Read 1 MB sequentially from memory       &250,000 ns & 0.25 ms\\
        Round trip within same datacenter (1 Gbps)&500,000 ns & 0.5 ms\\
        Read 1 MB sequentially from SSD          &1,000,000 ns & 1 ms\\
        Disk seek (fetch from new disk location) &10,000,000 ns & 10 ms\\
        Read 1 MB sequentially from disk         &20,000,000 ns & 20 ms\\
        Send packet CA $\to$ Netherlands $\to$ CA &150,000,000 ns & 150 ms \\ \hline
        \end{tabular}

    \end{table}
\section{Conclusion}
\label{sec:conclude}
In this paper, we argue that the biggest obstacles in the way of bringing big data for IoT from laboratory to industry are the noninvariant issues. We have revealed four common types of noninvariant issues, including context, identity, relationship, and behavior noninvariants; with concrete examples from Smart Grids. Based on this knowledge, we propose the Graph of Virtual Actors analytics architecture, which combines three essential ideas: virtual entities, micro-services architecture, and graph database. Besides solving the noninvariant issues, GoVA was also designed to be scalable in term of computation, storage, and programmability.
\section*{Acknowledgements}
We would like to show our gratitude to Christian Thun Eriksen, Lead Architect at eSmart Systems. He has provided insight and expertise that greatly assisted the study, although he may not agree with all of the interpretations/conclusions of this paper. The views and opinions expressed in this paper are solely our own and do not necessarily represent the views or opinions of eSmart Systems.


\ifCLASSOPTIONcaptionsoff
  \newpage
\fi



\bibliographystyle{IEEEtran}
\bibliography{references}

\begin{thebibliography}{10}
\providecommand{\url}[1]{#1}
\csname url@samestyle\endcsname
\providecommand{\newblock}{\relax}
\providecommand{\bibinfo}[2]{#2}
\providecommand{\BIBentrySTDinterwordspacing}{\spaceskip=0pt\relax}
\providecommand{\BIBentryALTinterwordstretchfactor}{4}
\providecommand{\BIBentryALTinterwordspacing}{\spaceskip=\fontdimen2\font plus
\BIBentryALTinterwordstretchfactor\fontdimen3\font minus
  \fontdimen4\font\relax}
\providecommand{\BIBforeignlanguage}[2]{{%
\expandafter\ifx\csname l@#1\endcsname\relax
\typeout{** WARNING: IEEEtran.bst: No hyphenation pattern has been}%
\typeout{** loaded for the language `#1'. Using the pattern for}%
\typeout{** the default language instead.}%
\else
\language=\csname l@#1\endcsname
\fi
#2}}
\providecommand{\BIBdecl}{\relax}
\BIBdecl

\bibitem{laney2001}
D.~Laney, ``3d data management: Controlling data volume, velocity, and
  variety,'' 2001.

\bibitem{gartner}
\BIBentryALTinterwordspacing
Gartner. (2012) Big data. [Online]. Available:
  \url{http://www.gartner.com/it-glossary/big-data}
\BIBentrySTDinterwordspacing

\bibitem{gartner2015culture}
\BIBentryALTinterwordspacing
G.~B. I. . A.~S. 2015. (2015) Gartner says business intelligence and analytics
  leaders must focus on mindsets and culture to kick start advanced analytics.
  [Online]. Available: \url{http://www.gartner.com/newsroom/id/3130017}
\BIBentrySTDinterwordspacing

\bibitem{lavalle2011big}
S.~LaValle, E.~Lesser, R.~Shockley, M.~S. Hopkins, and N.~Kruschwitz, ``Big
  data, analytics and the path from insights to value,'' \emph{MIT sloan
  management review}, vol.~52, no.~2, p.~21, 2011.

\bibitem{boyd2012critical}
D.~Boyd and K.~Crawford, ``Critical questions for big data: Provocations for a
  cultural, technological, and scholarly phenomenon,'' \emph{Information,
  communication \& society}, vol.~15, no.~5, pp. 662--679, 2012.

\bibitem{fan2013mining}
W.~Fan and A.~Bifet, ``Mining big data: current status, and forecast to the
  future,'' \emph{ACM SIGKDD Explorations Newsletter}, vol.~14, no.~2, pp.
  1--5, 2013.

\bibitem{ioannidis2005most}
J.~P. Ioannidis, ``Why most published research findings are false,'' \emph{PLoS
  Med}, vol.~2, no.~8, p. e124, 2005.

\bibitem{good2012common}
P.~I. Good and J.~W. Hardin, \emph{Common errors in statistics (and how to
  avoid them)}.\hskip 1em plus 0.5em minus 0.4em\relax John Wiley \& Sons,
  2012.

\bibitem{weisberg2011bias}
H.~I. Weisberg, \emph{Bias and causation: Models and judgment for valid
  comparisons}.\hskip 1em plus 0.5em minus 0.4em\relax John Wiley \& Sons,
  2011, vol. 885.

\bibitem{efron2012large}
B.~Efron, \emph{Large-scale inference: empirical Bayes methods for estimation,
  testing, and prediction}.\hskip 1em plus 0.5em minus 0.4em\relax Cambridge
  University Press, 2012, vol.~1.

\bibitem{walewski2013project}
J.~Walewski, M.~Bauer, N.~Bui, P.~Giacomin, N.~Gruschka, S.~Haller, E.~Ho,
  R.~Kernchen, M.~Lischka, J.~De~Loof \emph{et~al.}, ``Project deliverable d1.
  5--final architectural reference model for the iot v3.0,'' 2013.

\bibitem{gazis2013unified}
V.~Gazis, M.~Strohbach, N.~Akiva, and M.~Walther, ``A unified view on data path
  aspects for sensing applications at a smart city scale,'' in \emph{Advanced
  Information Networking and Applications Workshops (WAINA), 2013 27th
  International Conference on}.\hskip 1em plus 0.5em minus 0.4em\relax IEEE,
  2013, pp. 1283--1288.

\bibitem{strohbach2015towards}
M.~Strohbach, H.~Ziekow, V.~Gazis, and N.~Akiva, ``Towards a big data analytics
  framework for iot and smart city applications,'' in \emph{Modeling and
  Processing for Next-Generation Big-Data Technologies}.\hskip 1em plus 0.5em
  minus 0.4em\relax Springer, 2015, pp. 257--282.

\bibitem{lee2008cyber}
E.~A. Lee, ``Cyber physical systems: Design challenges,'' in \emph{2008 11th
  IEEE International Symposium on Object and Component-Oriented Real-Time
  Distributed Computing (ISORC)}.\hskip 1em plus 0.5em minus 0.4em\relax IEEE,
  2008, pp. 363--369.

\bibitem{popovicicapturing}
D.~Popovici and G.~Privat, ``Capturing the structure of internet of things
  systems with graph databases,'' 2015.

\bibitem{ibm2012temperature}
\BIBentryALTinterwordspacing
D.~Gibson. (2012) Is your big data hot, warm, or cold? [Online]. Available:
  \url{http://www.ibmbigdatahub.com/blog/your-big-data-hot-warm-or-cold}
\BIBentrySTDinterwordspacing

\bibitem{miller2013graph}
J.~J. Miller, ``Graph database applications and concepts with neo4j,'' in
  \emph{Proceedings of the Southern Association for Information Systems
  Conference, Atlanta, GA, USA}, vol. 2324, 2013.

\bibitem{xin2013graphx}
R.~S. Xin, J.~E. Gonzalez, M.~J. Franklin, and I.~Stoica, ``Graphx: A resilient
  distributed graph system on spark,'' in \emph{First International Workshop on
  Graph Data Management Experiences and Systems}.\hskip 1em plus 0.5em minus
  0.4em\relax ACM, 2013, p.~2.

\bibitem{shao2013trinity}
B.~Shao, H.~Wang, and Y.~Li, ``Trinity: A distributed graph engine on a memory
  cloud,'' in \emph{Proceedings of the 2013 ACM SIGMOD International Conference
  on Management of Data}.\hskip 1em plus 0.5em minus 0.4em\relax ACM, 2013, pp.
  505--516.

\bibitem{netflix2016}
\BIBentryALTinterwordspacing
T.~N.~T. Blog. (2016) Netflix data benchmark: Benchmarking cloud data stores.
  [Online]. Available: \url{http://techblog.netflix.com/}
\BIBentrySTDinterwordspacing

\bibitem{ebay2006}
R.~Shoup and D.~Pritchett, ``The ebay architecture,'' 2006.

\bibitem{amazon2007}
\BIBentryALTinterwordspacing
J.~Rohde. (2007) Amazon architecture. [Online]. Available:
  \url{http://highscalability.com/blog/2007/9/18/amazon-architecture.html}
\BIBentrySTDinterwordspacing

\bibitem{chris2015microservices}
\BIBentryALTinterwordspacing
C.~Richardson. (2015) Pattern: Microservices architecture. [Online]. Available:
  \url{http://microservices.io/patterns/microservices.html}
\BIBentrySTDinterwordspacing

\bibitem{abowd1999towards}
G.~D. Abowd, A.~K. Dey, P.~J. Brown, N.~Davies, M.~Smith, and P.~Steggles,
  ``Towards a better understanding of context and context-awareness,'' in
  \emph{International Symposium on Handheld and Ubiquitous Computing}.\hskip
  1em plus 0.5em minus 0.4em\relax Springer, 1999, pp. 304--307.

\bibitem{dang2015role}
T.-H. Dang-Ha, R.~Olsson, and H.~Wang, ``The role of big data on smart grid
  transition,'' in \emph{2015 IEEE International Conference on Smart
  City/SocialCom/SustainCom (SmartCity)}.\hskip 1em plus 0.5em minus
  0.4em\relax IEEE, 2015, pp. 33--39.

\bibitem{taohong2015whitepaper}
T.~Hong and M.~Shahidehpour, ``Load forecasting case study,'' National
  Association of Regulatory Utility Commissioners, Tech. Rep., 2015.

\bibitem{hong2014global}
T.~Hong, P.~Pinson, and S.~Fan, ``Global energy forecasting competition 2012,''
  \emph{International Journal of Forecasting}, vol.~30, no.~2, pp. 357--363,
  2014.

\bibitem{jonas2006identity}
J.~Jonas, ``Identity resolution: 23 years of practical experience and
  observations at scale,'' in \emph{Proceedings of the 2006 ACM SIGMOD
  international conference on Management of data}.\hskip 1em plus 0.5em minus
  0.4em\relax ACM, 2006, pp. 718--718.

\bibitem{Jonas2010}
------, ``Smart sensemaking systems, first and foremost, must be expert
  counting systems,'' \emph{International Risk Assessment and Horizon Scanning
  Symposium (IRAHSS)}, 2010.

\bibitem{qdr2006benefits}
U.~D. of~Energy., ``Benefits of demand response in electricity markets and
  recommendations for achieving them,'' 2006.

\bibitem{ananthanarayanan2011disk}
G.~Ananthanarayanan, A.~Ghodsi, S.~Shenker, and I.~Stoica, ``Disk-locality in
  datacenter computing considered irrelevant.'' in \emph{HotOS}, vol.~13, 2011,
  pp. 12--12.

\bibitem{dean2009designs}
J.~Dean, ``Designs, lessons and advice from building large distributed
  systems,'' \emph{Keynote from LADIS}, p.~1, 2009.

\end{thebibliography}

\end{document}